\documentclass[prc,aps,nofootinbib,showkeys,showpacs,twocolumn]{revtex4} 

\usepackage{amsmath}
\usepackage{amssymb}
\usepackage{amsfonts}
\usepackage{color}

\usepackage{epsfig}
\usepackage{graphicx}
\begin{document}
\date{\today}

\title{Symmetry breaking and fluctuations within stochastic mean-field dynamics: importance of initial quantum fluctuations}
 
\author{Denis Lacroix} \email{lacroix@ganil.fr}
\affiliation{GANIL, CEA and IN2P3, Bo\^ite Postale 5027, 14076 Caen Cedex, France}

\author{Sakir Ayik}
\affiliation{Physics Department, Tennessee Technological University, Cookeville, TN 38505, USA}

\author{Bulent Yilmaz}
\affiliation{Physics Department, Ankara University, Tandogan 06100, Ankara, Turkey}

\begin{abstract}  
Dynamics of spontaneous symmetry breaking and fluctuations in the Lipkin-Meshkov-Glick model are investigated in
a stochastic mean-field approach. Different from the standard mean-field, in the stochastic approach, 
initial state fluctuations, are incorporated. In weak coupling, the approach perfectly 
reproduces the exact quantal dynamics. On the other hand, for increasing coupling strength, above the symmetry 
breaking threshold, the approach provides description of gross properties (i.e. time averaged behavior) of the exact quantal evolution.  
\end{abstract}


\maketitle

The mean-field description of a many-body system, i.e. the Hartree-Fock (HF)
and/or time-dependent Hartree-Fock theory (TDHF), provides a simple tool for descriptions
of certain aspects of complex quantum systems \cite{Bla86}. For example, in the constrained
Hartee-Fock method, by explicitly breaking certain symmetries of the underlying Hamiltonian in 
static calculation, it is possible to describe topology of a quantum phase transitions \cite{Rin80}. 
However, it is well-known that  the mean-field approximation is suitable for the description 
of mean values of one-body observables, while quantum fluctuations of collective variables are 
severely underestimated \cite{Abe96}. Numerous approaches have been proposed either deterministic or 
stochastic to extended mean-field and describe fluctuations in collective space \cite{Lac04,Sim07}. 
Most often, these approaches are too complex to be applied in realistic situations
with actual computational power.  A second limitation of mean-field dynamics is that it can not 
describe spontaneous symmetry breaking during dynamical evolution. If certain symmetries are present 
in the initial state, these symmetries are preserved during the evolution  \cite{Bla86,Rin80}. 
Accordingly, mean-field cannot describe  
physical effects related to spontaneous symmetry breaking including molecule dissociation, spontaneous 
magnetization, and spontaneous fission in nuclei.  

Both dynamical symmetry breaking and lack of fluctuations are related to the absence of quantal effects 
in collective space and consequently collective motion appears nearly classical in the
mean-field dynamics. To overcome this difficulty, the mean-field approximation should be improved by
considering a more general wave function by coherent superposition of Slater determinants, 
such as in the time-dependent generator coordinate method \cite{Goe80,Goe81}. However, at present, applications 
of this method can be made only in a very restricted collective space and mostly along the adiabatic potential energy surface.
Here, we employ a stochastic approach,  which is much simpler than the generator coordinate method, and is based
on the fact that initial state fluctuations (quantal and thermal) dominate the fluctuation dynamics at low energies
\cite{Her84,Kay94}. This idea has been proposed nearly 30 years ago by Esbensen et al. in a macroscopic 
model of nuclear reactions \cite{Esb72,Das85}, and more recently tested in heavy-ion fusion reactions \cite{Ayi10}.  
Following a similar idea, recently, a stochastic mean-field approach (SMF) has been proposed \cite{Ayi08} to treat 
fluctuations beyond mean-field description. In the standard mean-field dynamics, ignoring quantal and thermal fluctuations, 
the initial state is specified in a deterministic manner: a given initial state leads to a well defined
final outcome. In the SMF, on the other hand, initial state fluctuations are incorporated in a stochastic approximation. 
Consequently, an ensemble of events are generated starting from a specified distribution of initial states. 
It is shown in ref. \cite{Ayi08}, in small amplitude limit, that
this  approach gives rise to the same expression for dispersions of one-body observables as the one obtained
in the variational description of Balian and V\'en\'eroni (BV) \cite{Bal81,Bal92}. In other applications, the average version 
of SMF theory was recently employed \cite{Ayi09,Was09,Yil11} to successfully reconcile onset of dissipation in TDHF 
and to calculate transport coefficients for relative momentum and nucleon-exchange in deep-inelastic heavy-ion collisions 
\cite{Randrup}. 

Recently, the variational approach of BV  has been  applied to nuclear reactions \cite{Bro08,Sim11}. Similarly to 
the standard mean-field description, the approach cannot describe spontaneous symmetry breaking mechanism, unless 
a symmetry-breaking density is used in the variational principle.  Therefore, the variational 
approach can only provide a poor approximation for dynamical evolution in the case of spontaneous symmetry breaking (SSB) 
(see Figure 6 of ref. \cite{Bon85}).  Currently, a realistic description of spontaneous symmetry breaking in the mean-field 
framework remains an open problem.  For this reason it is worthwhile to test whether the SMF approach  overcomes this difficulty. 
In the SMF approach the initial state is not the standard HF state,  but specified by a suitable distribution.
Even if the HF state respects a symmetry, in the SMF, this symmetry may be  broken  initially event by event. Consequently 
one might anticipate that, contrary to the original TDHF and/or BV methods, in the SMF approach it may be possible to  
treat the onset of the SSB.  We illustrate here that this is indeed the case. 

As a test case, the Lipkin-Meshkov-Glick (LMG) Model \cite{Lip65, Aga66, Sev06, Rin80} is considered here. This model consists of $N$ particles 
distributed in two N-fold degenerated single-particle states separated by an energy $\varepsilon$. The associated Hamiltonian 
is given by (taking $\hbar=1$),
\begin{eqnarray}
H = \varepsilon J_z - V(J^2_x  -  J_y^2) , 
\label{eq:hamillipkin}
\end{eqnarray}
where $V$ denotes the interaction strength while $J_i$ ($i=x$, $y$, $z$), are the quasi-spin operators defined as
\begin{eqnarray} 
J_z &=& \frac{1}{2} \sum_{p=1}^{N} \left(c^\dagger_{+,p}c_{+,p} - c^\dagger_{-,p}c_{-,p}\right) , \nonumber \\
J_x &=& \frac{1}{2} (J_+ + J_-), ~~~J_y = \frac{1}{2i} (J_+ - J_-)
\end{eqnarray}
with $J_+ = \sum_{p=1}^{N} c^\dagger_{+,p}c_{-,p}$, $J_- = J_+^\dagger$ and where 
$c^\dagger_{+,p}$ and $c^\dagger_{-,p}$ are creation operators associated with the upper and lower single-particle levels.
In the following, energies and times are given in $\varepsilon$ and $\hbar/\varepsilon$ units respectively.

This model has the advantage to be exactly solvable both in the static \cite{Rin80} and dynamical case \cite{Kan80,Kri77}. 
The LMG model is known to present a spontaneous symmetry breaking in mean-field theory 
as the interaction $V$ increases (see Figure \ref{fig1:SMFlipkin} below). Therefore, this model is perfectly suitable to investigate if the SMF approach is able to treat the SBB. 
 
The Hartree-Fock (or Mean-Field) solution is obtained by introducing 
the Slater Determinant trial states written as  $| \Phi 
\rangle = \Pi_{p=1}^N a^\dagger_{0,p} | - \rangle$, where the HF single-particle states are given by
\begin{eqnarray}
a^\dagger_{0,p} &=& \cos(\alpha) c^\dagger_{-,p} + \sin(\alpha) e^{i\varphi} c^\dagger_{+,p}.
\end{eqnarray}
The HF solution is obtained by minimizing  the mean-field energy with respect to variables $\alpha$
and $\varphi$,
\begin{eqnarray}
{\cal E}_{\rm HF}[\alpha , \varphi] &=& 
-\frac{\varepsilon N}{2} \left\{ \cos(2\alpha) + \frac{\chi}{2} \sin^2(2\alpha) \cos(2\varphi) \right\},
\label{eq:hflipkin}
\end{eqnarray}
where $\chi = V(N-1) / \varepsilon$. In Fig. \ref{fig1:SMFlipkin}, ${\cal E}_{\rm HF}[\alpha , 0] $ is shown for different $\chi$ parameters.
When the strength parameter is larger than  a critical value ($\chi > 1$), the parity symmetry is broken in $\alpha$ direction. 
\begin{figure}[htbp] 
\includegraphics[width=0.9\linewidth]{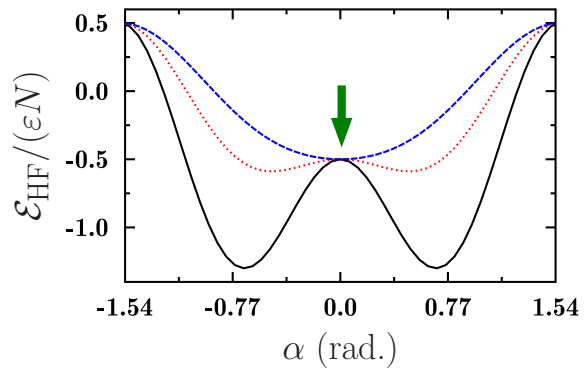} 
\caption{(color online). Evolution of the Hartree-Fock energy ${\cal E}_{\rm HF}$ as a function of $\alpha$ for $\chi=0.5$ (dashed line), 
$\chi=1.8$ (doted line) and $\chi=5$ (solid line) for $N=40$ particles.
The arrow indicates the initial condition used in the SMF dynamics.} 
\label{fig1:SMFlipkin} 
\end{figure} 

The mean-field evolution can be formulated either  in the Schr\"odinger \cite{Kri77} or Heisenberg picture. 
Here, we employ the second  option. We consider the expectation values of the quasi-spin operators $j_i \equiv 
\langle J_i \rangle/N$ (for $i=x$,  $y$ and $z$). In the mean-field approximation, it is possible to derive a set of coupled equations for the expectation values of the quasi-spin operators as,
\begin{eqnarray}
\frac{d}{dt} 
\begin{pmatrix}
 j_x 
    \\
  j_y  \\
 j_z 
\end{pmatrix}
&=& \varepsilon 
\begin{pmatrix}
 0   & -1 + \chi  j_z  &  \chi   j_y  \\
 1+ \chi   j_z    &  0 & \chi  j_x  \\
 -2 \chi   j_y & -2 \chi   j_x  & 0
\end{pmatrix}
\begin{pmatrix}
 j_x 
    \\
 j_y \\
 j_z 
\end{pmatrix}.
\label{eq:mfspin}
\end{eqnarray}
Initially, we prepare the system in the state $| j, -j \rangle$, i.e. $\alpha=0$, which means that all particles are placed in the
lower single-particle states. This case is indicated  by an arrow in Figure \ref{fig1:SMFlipkin}. In this state,
initial expectation values of quasi-spin components are  $j_z(t_0)= -1/2$, $j_x(t_0)=j_y(t_0) =0$. This 
initial condition is a stationary solution of Eq. (\ref{eq:mfspin}). When the strength parameter is larger that critical value
$\chi>1$, the initial state is at the saddle point. Since mean-field cannot break the symmetry, the system will remain 
at the saddle point. Therefore, it is not possible to describe
onset of SSB in the standard mean-field framework. This situation is similar to the classical object positioned at $\alpha=0$.
In exact quantal description, since the initial state is not an eigenstate  
of the Hamiltonian $H$, different spin components and their correlations change in time. The difference between the exact and 
the mean-field evolution is that quantum fluctuations are properly taken into account in the exact evolution.

In the SMF approach, the expectation values of the quasi-spin operators obey the same set of equations given by Eq. (\ref{eq:mfspin}),
except that the initial conditions are different. In order to simulate quantum fluctuations in an approximate manner, in the SMF approach
\cite{Ayi08}, an initial ensemble of single-particle density matrices is prepared around the same state $| j, -j \rangle$ used in 
the exact evolution.  According to the stochastic properties of the initial state, it is possible to  determine the initial 
distributions of expectation values of quasi-spin operators. We find that that the $z$ quasi-spin component is not a fluctuating quantity with a mean value $j_z(t_0) = -\frac{1}{2} $.
On the other hand, the $x$ and $y$ quasi-spin components are uncorrelated Gaussian random numbers with zero mean values, 
\begin{eqnarray}
\overline{j^\lambda_x(t_0)}=\overline{j^\lambda_y(t_0)} = 0,
\end{eqnarray}
and second moments determined by, 
\begin{eqnarray}
\overline{j^\lambda_x(t_0)j^\lambda_x(t_0)} = \overline{j^\lambda_y(t_0) j^\lambda_y(t_0)} = \frac{1}{4N}. 
\end{eqnarray}
We note that even if all trajectories start from the top of the energy landscape (the arrow in Fig. \ref{fig1:SMFlipkin}) and the
system has good parity in average, in the SMF evolution this symmetry is broken event by event due to 
non-zero values of the spin components along the $x$ and $y$ axis. 

In the SMF, mean values and fluctuations of observables are obtained by performing average of expectation values over the
generated ensemble. The mean  values of the quasi-spin components  and associated dispersions, denoted respectively by
$J_i$ and $\Delta^2_i $, are given by
\begin{eqnarray}
J_i(t)  & = & N \overline{j^\lambda_i(t)}, ~~~
\Delta^2_i(t) =  N^2  \left(\overline{{(j^{\lambda }_i)}^2 } - {(\overline{j^\lambda_i})}^2 \right).
\end{eqnarray}

\begin{figure}[htbp] 
\begin{center}
\includegraphics[width=0.9\linewidth]{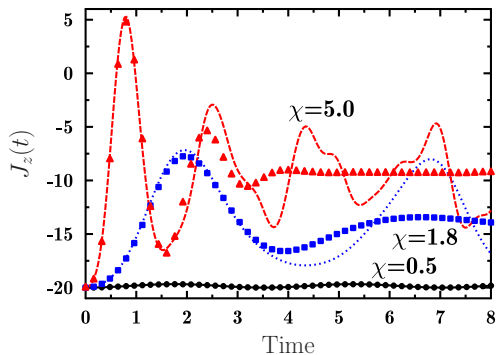} 
\end{center}
\caption{(color online) Exact evolution of the $z$ quasi-spin component obtained when the initial state is $|j,-j\rangle$ for three 
different values of $\chi$: $\chi = 0.5$ (solid line), $\chi=1.8$ (dotted line) and $\chi=5.0$ (dashed line) for $N=40$ particles. The corresponding 
results obtained with the SMF simulations are shown with circles, squares and triangles respectively. } 
\label{fig2:SMFlipkin} 
\end{figure} 

\begin{figure}[htbp] 
\begin{center}
\includegraphics[width=0.8\linewidth]{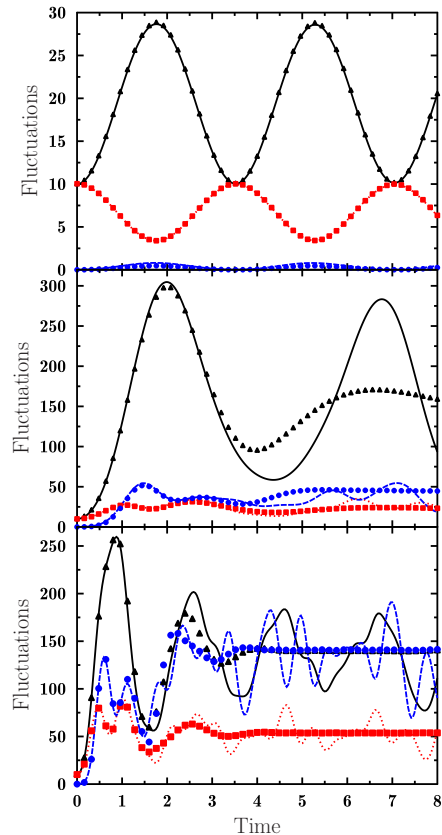} 
\end{center}
\caption{(color online)  Exact evolution of dispersions of quasi-spin operators obtained when the initial state is $|j,-j\rangle$ for three 
different values of $\chi$, from top to bottom $\chi = 0.5$ (top), $\chi=1.8$ (middle) and $\chi=5.0$ (bottom) are shown. In each case, 
solid, dashed and dotted lines correspond to $\Delta^2_x (t)$, $\Delta^2_y (t)$ and $\Delta^2_z (t)$, respectively.
In each case, results of the SMF simulations are shown with triangles ($\Delta^2_x$), squares ($\Delta^2_y$) and 
circles ($\Delta^2_z$). } 
\label{fig3:SMFlipkin} 
\end{figure} 

In Fig. \ref{fig2:SMFlipkin}, the exact and SMF evolutions of the $z$ quasi-spin component obtained when the initial state is $|j,-j\rangle$ 
for three different values of $\chi$: $\chi = 0.5$, $\chi=1.8$ and $\chi=5.0$ are shown. In TDHF, for 40 particles, 
this component remains constant and equal to -20. In both the exact results and the  SMF simulations, 
mean values of $x$ and $y$ components are zero. In this and following figure, the SMF simulation are carried out 
using a set of $10^5$ trajectories. Simulations are performed using Runge-Kutta or order 2 algorithm with a time step of $0.01$.

The evolutions of dispersions of quasi-spin components obtained in the SMF simulations are shown in Fig. \ref{fig3:SMFlipkin}, and
compared with the exact results, $\Delta^2_i (t) = \langle J^2_i \rangle - \langle J_i \rangle^2$, obtained starting from the state 
$|j,-j\rangle$.  We note that, in the standard TDHF dynamics, since the state  $|j,-j\rangle$, does not evolve in time, dispersions of the 
quasi-spin variables remain constant and  equal to their initial values, $\Delta^2_x = \Delta^2_y = \frac{N}{4} $ and $\Delta^2_z = 0$. 
Below the critical value of the strength parameter ($\chi=0.5$), where energy can be regarded as nearly harmonic around $\alpha=0$ (see Figure \ref{fig1:SMFlipkin}), results obtained in the SMF simulations can hardly be distinguished from the exact solution. Only a small difference 
is noticeable in the $z$ component. A similar result is obtained in ref. \cite{Bon85} with the BV description. The fact that both approaches
produce very similar results in the harmonic limit is not surprising, since it was shown that they both contain the same physical
information in this limit \cite{Ayi08}. Above the critical strength ($\chi=1.8$ and $5$) 
(middle and bottom panel of Figure \ref{fig3:SMFlipkin}), the BV description has been shown  to lead to very bad results \cite{Bon85},
when calculations start from the same initial condition. Here, we see that the SMF approach provides a 
fairly good reproduction of gross properties of the exact dynamics. In particular, during the early times, the SMF simulations can not be
distinguished from the exact evolution. During the long time evolution, the SMF simulations describe time-averaged behavior of the exact 
dynamics very well.  As seen from Fig. \ref{fig2:SMFlipkin}, a similar agreement is obtained for the mean value of the $z$ component of
quasi-spin for all values of the strength parameter $\chi$. The present example, clearly demonstrates the ability of the SMF approach to describe gross properties of mean values and fluctuations for any strength of the interaction.   

It is well known that the standard mean-field theory provides a poor description for fluctuations of collective motion, and it essentially 
treats the collective motion in a classical approximation. The SMF approach makes an attempt to correct this shortcoming by incorporating quantal and thermal fluctuations in the initial state. 
In this work, we test the approach in the LMG model.  
As seen in Fig. \ref{fig2:SMFlipkin} and Fig. \ref{fig3:SMFlipkin}, 
the SMF simulations provide nearly perfect description for non-trivial oscillations during early evolution of mean values and dispersions of quasi-spin operators. Over the long time interval, simulations also provide a satisfactory description for the gross properties, i.e., time averaged behavior of the mean values and dispersions of quasi-spin operators. We should note that we do not expect that such a simple SMF approach provides a detailed quantum mechanical feature of the evolution. In particular, possible 
interferences between different trajectories are neglected as well as possible tunneling effects. 
Nevertheless, the stochastic method presented here provides a suitable framework beyond mean-field for describing dynamics of fluctuations and for understanding spontaneous symmetry breaking in complex quantum systems from a quasi-classical perspective.   

\section*{Acknowledgment}  
S. A. gratefully acknowledges France-US Institute for Physics of Exotic Nucleus for financial support. S. A. and B. Y. gratefully acknowledge GANIL for hospitality extended to them during their visit. B. Y. gratefully  acknowledges TUBITAK for financial support. This work is supported in part by the US DOE Grant No. DE-FG05-89ER40530.


\begin{thebibliography} {99}
\bibitem{Bla86} J. P. Blaizot and G. Ripka, {\it Quantum Theory of Finite Systems}, (MIT Press, Cambridge,  
Massachusetts, 1986).  
\bibitem{Rin80} P. Ring and P. Schuck, {\it The Nuclear Many-Body Problem}
(Springer-Verlag, New-York, 1980).

\bibitem{Abe96}Y. Abe, S. Ayik, P.-G. Reinhard, E. Suraud, Phys. Rep. {\bf 275}, 49 (1996).

\bibitem{Lac04} D. Lacroix, S. Ayik, and Ph. Chomaz, Prog. Part. Nucl. Phys.
{\bf 52}, 497 (2004).
\bibitem{Sim07} C. Simenel, B. Avez, and D. Lacroix, in Lecture Notes
of the International Joliot-Curie School, (2007), arXiv:0806.2714.

\bibitem{Goe80} K. Goeke, and P. G. Reinhard, Ann. Phys. (NY) {\bf 124}, 249 (1980).
\bibitem{Goe81} K. Goeke, P. G. Reinhard and D. J. Rowe, Nucl. Phys.  {\bf A359}, 408 (1981). 

\bibitem{Her84} M. F. Herman and E. Kluk, Chem. Phys. {\bf 91}, 27 (1984).

\bibitem{Kay94} K. G. Kay, J. Chem. Phys. {\bf 100}, 4432 (1994). {\it ibid}, J. Chem. Phys. {\bf 101}, 2250 (1994). 

\bibitem{Esb72} H. Esbensen, A.Winther, R. A. Broglia, and C. H. Dasso, Phys.
Rev. Lett. {\bf 41}, 296 (1978).

\bibitem{Das85} C. H. Dasso, in {\it Proceedings of the la Rabida Int. Summer School
on Nuclear Physics}, edited by M. Lozano and G. Madurga,
(World Scientific, Singapore, 1985), p. 398.

\bibitem{Ayi10} S. Ayik, B. Yilmaz and D. Lacroix, Phys. Rev. {\bf C81}, 034605 (2010).


\bibitem{Ayi08} S. Ayik, Phys. Lett. {\bf B658}, 174 (2008). 


\bibitem{Bal81} R. Balian and M. V\'en\'eroni, Phys. Rev. Lett. {\bf 47}, 1353
(1981).
\bibitem{Bal92} R. Balian and M. Ve\'n\'eroni, Ann. Phys. (N.Y.) {\bf 216}, 351
(1992).

\bibitem{Was09} K. Washiyama, S. Ayik, and D. Lacroix, Phys. Rev. {\bf C80},
031602(R) (2009).

\bibitem{Yil11} B. Yilmaz, S. Ayik, D. Lacroix, and K. Washiyama, Phys. Rev.
{\bf C83}, 064615 (2011).

\bibitem{Ayi09} S. Ayik, K. Washiyama, and D. Lacroix, Phys. Rev. {\bf C79}, 054606
(2009).

\bibitem{Randrup} J. Randrup, Nucl. Phys. {\bf A 307}, 319 (1978); {\it ibid} Nucl. Phys. {\bf A327}, 490 (1979);
383, 468 (1982).

\bibitem{Bro08} J. M. A. Broomfield and P. D. Stevenson, J. Phys. {\bf G35}, 095102
(2008).

\bibitem{Sim11} C. Simenel, Phys. Rev. Lett. {\bf 106}, 112502 (2011).


\bibitem{Bon85} P. Bonche and H. Flocard, Nucl. Phys. {\bf A437},  189 (1985).



\bibitem{Lip65} H. J. Lipkin, N. Meshkov and A. J. Glick, Nucl. Phys. {\bf A62}, 188 (1965).
\bibitem{Aga66} N. M. D. Agassi and H. J. Lipkin, Nucl. Phys. {\bf A86}, 321 (1966).
\bibitem{Sev06} A. P. Severyukhin, M. Bender, and P.-H. Heenen, Phys. Rev. 
{\bf C74}, 024311 (2006).
\bibitem{Kan80} K.-K. Kan, P. C. Lichtner, M. Dworzecka and J. J. Griffin, Phys. Rev. {\bf C21}, 1098 (1980).
\bibitem{Kri77} S. J. Krieger, Nucl. Phys. {\bf A276}, 12 (1977). 
\end{thebibliography}
\end{document}